\documentclass[11pt]{article}
\usepackage{amsmath,amssymb}
\usepackage{amsfonts}
\usepackage[dvipdfm]{graphicx}
\usepackage[dvipdfmx,bookmarksnumbered=true,breaklinks=true]{hyperref}
\usepackage{fancyhdr}

\newcommand{\etal}{{\it et al.}}
\newcommand{\eqnref}[1]{Eqn.~(\ref{#1})}		
\newcommand{\ri}{{\rm i}}						

\renewcommand{\Xi}{\Xi}

\newcommand{\wsq}{\widetilde{\square}}

\def\be{\begin{equation}}
\def\ee{\end{equation}}
\def\bea{\begin{eqnarray}}
\def\eea{\end{eqnarray}}

\def\bb{\begin{equation*}}
\def\eb{\end{equation*}}
\def\beb{\begin{eqnarray*}}
\def\eeb{\end{eqnarray*}}

\begin{document}

\thispagestyle{empty}
\begin{titlepage}

\begin{flushright}
TUW-10-08
\end{flushright}

\vspace{0.3cm}
\boldmath
\begin{flushleft}
  {\large {\bf On NCQFT and dimensionless insertions}}
\end{flushleft}
\unboldmath
\vspace{0.5cm}
\begin{flushleft}
{\bf
Manfred Schweda\footnote{mschweda@tph.tuwien.ac.at}
and 
Michael~Wohlgenannt\footnote{michael.wohlgenannt@tuwien.ac.at}
}                                                                                                                                         
\end{flushleft}
\begin{flushleft}
{\it
Institute for Theoretical Physics, Vienna University of Technology,\\
Wiedner Hauptstrasse 8-10, A-1040 Vienna, Austria
}
\end{flushleft}
\vskip 1em

\vfill
\begin{abstract}
\noindent
In these notes, we aim at a precise definition of the tree level action for the noncommutative scalar and gauge field theories on four-dimensional canonically deformed Euclidean space. As tools to achieve this goal we employ power counting and normalization conditions.
\end{abstract}
PACS: 11.10.Nx, 11.15.-q\\
Keywords: NCQFT, normalization conditions, dimensionless insertions

\end{titlepage}

\section{Introduction}

In this short letter, we discuss the precise definition of translation invariant models for noncommutative field theories at tree level. This is not a trivial problem due to the presence of the deformation parameter $\theta_{\mu\nu}$, which has mass dimension $-2$. In the past, noncommutative quantum field theories have been constructed by naively replacing any product of fields by a star-product, e.g. the Weyl-Moyal product
\be
\label{weyl-moyal}
\phi_a \star \phi_b \,(x) = e^{\frac i2 \theta_{\mu\nu} \partial^x_\mu
\partial_\nu^y} \phi_a(x) \phi_b(y)\Big|_{y\to x} \ne \phi_b \star \phi_a\,
(x)\,,
\ee
where $\phi_a$ stands for all possible fields contained in the model. 

At the classical level, the commutative action for scalar $\phi^4$ theory in four-dimensional Euclidean space is given by
\be
\Gamma^{(0)} = \int d^4x \left(
\frac 12 \partial_\mu \phi \, \partial_\mu \phi + \frac{m^2}2 \phi^2 + \frac\lambda{4!} \phi^4
\right)\,.
\ee
According to the above mentioned recipe, one obtains for the noncommutative counterpart 
\bea
\nonumber
\Gamma^{(0)} & = & \int d^4x \left(
\frac 12 \partial_\mu \phi \star \partial_\mu \phi + \frac{m^2}2 \phi \star \phi + \frac\lambda{4!} \phi^{\star 4}
\right)\\
\label{nc-naive}
& = & \int d^4x \left(
\frac 12 \partial_\mu \phi \, \partial_\mu \phi + \frac{m^2}2 \phi^2 + \frac\lambda{4!} \phi^{\star 4}
\right)
\,.
\eea
However, this action leads to the infamous UV/IR mixing property, which is an obstacle for general renormalizability. Among the possibilities to avoid this problem and to make sense of the perturbative expansion, we want to focus on the one suggested by Gurau {\it et al.}~\cite{Rivasseau:2008a}\footnote{Here, we want to consider translation invariant models only. Another modification leading to a renormalizable model has been discussed e.g. in \cite{Grosse:2004b}}. They modified the bi-linear part of the classical action in the following way:
\be
\label{gurau}
\Gamma^{(0)} = \int d^4 x \left(
\frac 12 \partial_\mu \phi \, \partial_\mu \phi + \frac {m^2}2 \phi^2 + \phi \frac 1{\wsq} \phi + \frac\lambda{4!} \phi^{\star 4}
\right)\,,
\ee
where we have introduced the notation $\tilde \partial_\mu = \theta_{\mu\nu}\partial_\nu$ and
$\wsq = \tilde \partial_\mu \tilde \partial_\mu$. We assume that $\theta_{\mu\nu}$ has full rank and choose the following representation
\be
\theta_{\mu\nu} = \theta \, \left(
\begin{array}{cccc} 0&1&& \\ -1&0&& \\ &&0&1 \\ &&-1&0 \end{array}
\right)\,.
\ee
Thus, we have $\wsq = \theta^2\, \Box$, with mass dimension $dim\,\wsq = -2$. Due to the modification of Gurau {\etal} the propagator in momentum space is given by 
\be
\label{prop}
\Delta(k) = \frac 1 { k^2 + m^2 + \frac1{\theta^2\, k^2} }\,.
\ee
As a consequence, the high momentum behaviour remains unchanged, but the non-perturbative region (for small k-values) changes dramatically and avoids UV/IR mixing \cite{Rivasseau:2008a,Blaschke:2008b}. This modifications can be implemented by taking advantage of the fact that the non-commutativity parameter introduces a length scale and operators of the form $\wsq \Box$ or $D^2 \tilde D^2$, respectively, where $D$ denotes the covariant derivative, have mass dimension zero. But clearly, one needs to distinguish between mass dimension and UV or scaling dimension. The latter effects the power counting, as has also been emphasized in \cite{Vilar:2009}.\\ 
In Section 2, we will discuss the general action for scalar fields and how to restrict it to \eqnref{gurau} using power counting arguments and normalization conditions. In Section 3, we will examine the tree-level action of noncommutative $U(1)$ gauge theory. Here, we discuss a gauge invariant implementation of the damping behaviour necessary in order to avoid UV/IR mixing; see e.g. \cite{Blaschke:2008a,Blaschke:2009a,Blaschke:2009b,Blaschke:2009d,Vilar:2009}. However, there are also different ways of implementing IR modifications of the propagator. The probably most promising one \cite{Blaschke:2009e} consistently implements the IR damping behaviour in the so-called soft breaking term - a method which is well known from the Gribov-Zwanziger approach to QCD \cite{Gribov:1978,Zwanziger:1989,Zwanziger:1993}.

\section{Scalar theory}

Initially, the whole community has used the action (\ref{nc-naive}). Motivated by renormalizability considerations, it was replaced by an improved version given in Eq.~(\ref{gurau}). Certainly, this is not the most general deformed extension one can think of.
Due to the fact that the non-commutativity described by $\theta^{\mu\nu}$ introduces a scale, and $\Box \wsq$ is a therefore dimensionless differential operator, the following general action is possible\footnote{There are also other possible terms, such as $\int d^4x\, \phi\star (\partial_\mu\phi)\star \phi \star (\tilde \partial_\mu\phi)$ and the other feasible combinations for the self-interaction.}:
\bea
\label{extension}
\Gamma^{(0)} & = & \int d^4 x \left(
\frac 12 \partial_\mu \phi (\Box \wsq)^{\alpha_1} \partial_\mu \phi + m^2 \phi (\Box \wsq)^{\alpha_2} \phi + \phi \frac
{(\Box \wsq)^{\alpha_3}}{\wsq} \phi \right.\\
\nonumber
&& \hspace{.8cm}
\left. + \frac 1{4!} \sum_i \lambda_i (\phi\star \phi) (\Box \wsq)^{n_i} (\phi\star \phi)
\right)\,,
\eea
where $\alpha_1,\alpha_2,\alpha_3$ and $n_i$ are arbitrary constants, and we have restricted ourselves to a Lagrangian of mass dimension 4. 
In order to finally obtain the action \eqref{gurau}, we have to postulate corresponding normalization conditions fixing the bi-linear part which is responsible for the desired propagator. The self-interaction terms are determined by power counting arguments and a corresponding normalization condition.

In momentum space, we postulate the following normalization conditions for the 2-point vertex function at tree-level, cf. also \cite{Tanasa:2008a}:
\bea
\label{AI}
k^2 \tilde \Gamma^{(0)}_2 (k) \Big|_{k^2=0} & = & \frac 1{\theta^2},\\
\label{AII}
\left( \tilde \Gamma^{(0)}_2 (k) - \frac 1{\theta^2k^2} \right)\Big|_{k^2=0} & = & m^2_{\textrm{phys}},\\
\label{AIII}
\frac d{dk^2} \left( \tilde \Gamma^{(0)}_2 (k) - \frac 1{\theta^2k^2} \right)\Big|_{k^2=0} & = & 1\,.
\eea
These three conditions define the form of the bi-linear part and lead to the propagator \eqref{prop}. These conditions fix the constants to $\alpha_1=\alpha_2=\alpha_3=0$ as the only consistent choice.

In a next step, we want to discuss the interaction terms which are governed by demanding power counting rernormalizability and locality (except for the star products). First of all, this assures that $n_i\ge 0$, $\forall i$. Since $\Box \wsq = \theta^2 \partial^4$, one has in momentum space $\theta^2k^4$. Thus, the corresponding interaction vertices contain additional factors $k^{\delta_i}$, where $\delta_i$ denotes the number of derivatives of the respective vertex. These factors have an important influence on the power counting.

One can derive with the same topological formulae as in the commutative case defining a Feynman graph $\gamma$ that the superficial degree of divergence is given by
\be
\label{powercounting}
D(\gamma) = 4 - E + \sum_i \delta_i\,,
\ee
with $\delta_i=4 n_i$, and 
where $E$ denotes the number of external legs of the graph $\gamma$. Thus, we see that the term $(\phi\star \phi) (\theta^2k^4)^{n_i}(\phi\star\phi)$ destroys power counting renormalizability for $n_i\ne 0$. Hence, we obtain $\delta_i=0$ and finally the following normalization condition
\bea
\tilde \Gamma^{(0)}_4(k_1,k_2,k_3,k_4)\Big|_{k_i=p_i} & = & (2\pi)^4 \frac \lambda 3 \left(
\cos \frac{p_1\tilde p_2}2 \cos \frac{p_3\tilde p_4}2
\right.\\
\nonumber
&& \hspace{-1cm}
\left. + \cos \frac{p_1\tilde p_3}2 \cos \frac{p_2\tilde p_4}2 + \cos \frac{p_1\tilde p_4}2\cos \frac{p_2\tilde p_3}2
\right)\,.
\eea
The action \eqref{extension} reduces to
\bb
\Gamma^{(0)} = \int d^4 x \left(
\frac 12 \partial_\mu \phi \, \partial_\mu \phi + \frac {m^2}2 \phi^2 + \phi \frac 1{\wsq} \phi + \frac\lambda{4!} \phi^{\star 4}
\right)\,,
\eb
and we have showed that also the interaction terms with additional derivatives do not appear. The same holds true even in the commutative model. However, one has to remark that derivatives in the usual $\phi^4$ model are forbidden by dimensional reasons, unless a scale is present in the theory.

In \cite{Blaschke:2008b}, it has been shown explicitly that the normalization conditions are preserved at one loop order. The renormalized propagator is given by
\be
\Delta'(p) = \frac Z { p^2 +m_{\textrm{phys}}^2 + \frac{a_{\textrm{phys}}^2}{p^2} + f(p^2) }\,,
\ee
where
\bea
Z & \equiv & 1 + \lambda \alpha \theta^2,\\
m_{\textrm{phys}}^2 & \equiv & m^2 + \frac{\lambda}{3(4\pi)^2} \left(
4 \Lambda^2 + m^2 \ln \left( \frac 1{\Lambda^2} \sqrt{\frac{m^4}4 - M^4} \right)
\right),\\
a_{\textrm{r}}^2 & \equiv & a^2 + \lambda \left(\frac 2{3(4 \pi\, \theta)^2} + \alpha a^2 \theta^2 \right) 
\,,
\eea
with UV cut-off $\Lambda$. The function $f(p^2)$ and  $\alpha\in \mathbb R$ are given in \cite{Blaschke:2008b}, 
$$
f(p^2) = \frac \lambda {6(4\pi)^2} m^2 \ln(\theta^2p^2) 
\,.
$$
Apart from the explicit momentum dependence, it seems more natural to define 
$$
a_\textrm{phys}(p^2) = a_{\textrm{r}} + p^2 f(p^2)\,.
$$
Although $a_\textrm{phys}$ now depends on the momentum, $p^2 f(p^2)\to 0$ at the renormalization point $p^2=0$. Therefore, it can 
very well be absorbed in the parameter $a$. By the usual wave function renormalization we can remove Z,
\be
\phi_{\textrm{phys}}  = Z^{-1/2} \phi\,.
\ee
For vanishing $\theta$, we of course obtain $Z=1$. At one-loop level, one then arrives at
\be
\tilde \Gamma^{(1)}_{2,\textrm{phys}} = k^2 + m_{\textrm{phys}}^2 + \frac{a_{\textrm{phys}}^2}{k^2}\,. 
\ee

As a summary, one can state that the action of a noncommutative scalar quantum field theory is defined in such a way that the bi-linear part may contain a non-local piece $\frac 12 \phi \frac 1{\wsq} \phi$ in order to obtain an IR damping for small $k$ values - whereas the interaction terms are determined by the requirement of power counting renormalizability.

\section{$U(1)$ Gauge theory}

Due to the tensor structure in a $U(1)$ gauge theory, there are many possibilities to construct an action at tree-level invariant under the noncommutative gauge transformation 
$$
\delta A_\mu = \partial_\mu \lambda - \ri [A_\mu\stackrel{\star}{,} \lambda] = D_\mu \lambda\,.
$$
In its general form, it reads
\bea
\label{extended-gauge}
\Gamma^{(0)} & = & \int d^4x \Bigg(
\frac 14 F_{\mu\nu} ( 1 + (\theta^2 D^4)^{m_1} )^{n_1} F_{\mu\nu} \\
\nonumber
&& \hspace{-1cm}
+ \frac \alpha 4 F_{\mu\nu} ( 1 + (\theta^2 D^4)^{m_2} )^{n_2} 
\frac 1{\theta^2 D^4} F_{\mu\nu} + \frac \beta 4 \tilde F ( 1 + (\theta^2 D^4)^{m_3} )^{n_3} \frac 1{\theta^2 D^4} \tilde F 
\Bigg)\,,
\eea
where $m_i,\, n_i$, $\alpha$ and $\beta$ are arbitrary constants. For simplicity, we choose $m_1=m_2=m_3=m$, $n_1=n_2=n_3=n$. The noncommutative field strength is given by
$$
F_{\mu\nu} = \partial_\mu A_\nu - \partial_\nu A_\mu - \ri [A_\mu \stackrel{\star}{,} A_\nu]\,,
$$
covariant derivatives are denoted by $D_\mu = \partial_\mu - \ri [A_\mu \stackrel{\star}{,} \cdot]$, and $\tilde F = \theta_{\mu\nu} F_{\mu\nu}$.

In order to calculate the propagator, we only consider the bi-linear part of the above action:
\bea
\label{bi-linear}
\Gamma^{(0)}_{bi} & = & \int d^4x \Bigg(
\frac 12 (\partial_\mu A_\nu - \partial_\nu A_\mu) ( 1 + (\theta^2 \partial^4)^m )^n \partial_\mu A_\nu \\
\nonumber
&& + \frac \alpha 2 (\partial_\mu A_\nu - \partial_\nu A_\mu) ( 1 + (\theta^2 \partial^4)^m )^n \frac 1{\theta^2 \partial^4} \partial_\mu A_\nu\\
\nonumber
&& +  \beta \tilde \partial_\mu A_\mu ( 1 + (\theta^2 \partial^4)^m )^n \frac 1{\theta^2 \partial^4} \tilde \partial_\nu A_\nu
\Bigg)\,.
\eea
This leads to the following expression for the vacuum polarization at tree level
\bea
\Pi_{\mu\nu}(k) & = & (g_{\mu\nu} - \frac{k_\mu k_\nu}{k^2}) ( 1+(\theta^2 k^4)^m)^n \left( k^2 + \frac \alpha {\theta^2 k^2} \right)
\\
\nonumber
&& \hspace{.8cm}
+ \beta (1+(\theta^2 k^4)^m)^n \frac {\tilde k_\mu \tilde k_\nu}{\theta^4 k^4}
\\
\nonumber
& \equiv & \Pi_1(k^2, \theta) \left( g_{\mu\nu} - \frac{k_\mu k_\nu}{k^2} \right) + \Pi_2(k^2,\theta) \tilde k_\mu \tilde k_\nu\,.
\eea
Including the gauge fixing and external sources, the bi-linear action (\ref{bi-linear}) reads
\bea
\Gamma^{(0)}_{bi} & = & \int d^4x \Bigg(
\frac 12 (\partial_\mu A_\nu - \partial_\nu A_\mu) ( 1 + (\theta^2 \partial^4)^m )^n \partial_\mu A_\nu \\
\nonumber
&& + \, \frac \alpha 2 (\partial_\mu A_\nu - \partial_\nu A_\mu) ( 1 + (\theta^2 \partial^4)^m )^n \frac 1{\theta^2 \partial^4} \partial_\mu A_\nu\\
\nonumber
&& + \, \beta \tilde \partial_\mu A_\mu ( 1 + (\theta^2 \partial^4)^m )^n \frac 1{\theta^2 \partial^4} \tilde \partial_\nu A_\nu\\
\nonumber
&& + \, B \left(  1 + (\theta^2 \partial^4)^m ) \right)^n \partial_\mu A_\mu + j_\mu A_\mu + Bj
\Bigg)\,.
\eea
In order to compute the propagator, we have to express $A_\mu$ in terms of the external sources:
\bea
A_\rho & = & (1+(\theta^2 \partial^4)^m)^{-n} (1+\frac \alpha{\theta^2\partial^4} )^{-1} \frac 1{\partial^2} 
\Bigg( j_\rho - \frac{\partial_\rho \partial_\sigma}{\partial^2} j_\sigma \\
\nonumber
&& \hspace{1.2cm}
- \frac {2\beta}{\theta^4 \partial^4(1+\frac 1{\theta^2\partial^4}(\alpha + 2\beta))} \frac{\tilde \partial_\rho \tilde \partial_\sigma}{\partial^2} j_\sigma 
\Bigg)
\eea
and, therefore, 
\bea
\widetilde {\frac {\delta A_\rho}{\delta j_\sigma}} & = & \frac 1{ (1+(\theta^2k^4)^m)^n(1+\frac \alpha{\theta^2k^4}) } \frac 1{k^2}
\Bigg( g_{\rho\sigma} - \frac{k_\rho k_\sigma}{k^2}\\
\nonumber
&& \hspace{1.3cm}
 - \frac{2\beta}{\theta^2 k^4 (1+\frac 1{\theta^2k^4}(\alpha + 2\beta)) }
\frac {\tilde k_\rho \tilde k_\sigma}{\tilde k^2}
\Bigg)\, .
\eea
In order to obtain the desired propagator
\be
\label{tree}
\Delta_{\mu\nu} (k) = \frac 1{k^2 + \frac 1{\tilde k^2}} \left( g_{\mu\nu} - \frac {k_\mu k_\nu}{k^2} \right)\,,
\ee
one has to choose 
\be
\label{choice1}
\{\alpha=1,\beta=0,m=n=0\}\,,
\ee 
or 
\be
\label{choice2}
\{\alpha=\beta=0,m=-1, n=1\}\,.
\ee
Choosing the proper values for the parameters corresponds to applying normalization conditions defining the tree-level propagator (\ref{tree}):
\bea
k^2 \Pi_1 \Big|_{k^2=0} & = & \frac 1{\theta^2} \,,\\
\left( \Pi_1 - \frac 1{\theta^2 k^2} \right) \Big|_{k^2=0} & = & 0\,, \\
\frac d{dk^2} \left( \Pi_1 - \frac 1{\theta^2 k^2} \right) \Big|_{k^2=0} & = & 1\,,\\
\Pi_2 \Big|_{k^2=0} & = & 0\,.
\eea

Having discussed the bi-linear parts of the action via normalization conditions, it is now straight forward to find also normalization conditions for the interaction terms of the action in order to define the total action at tree level. The resulting action reads
\be
\Gamma^{(0)} = \int d^4x \left(
\frac 14 F_{\mu\nu} F_{\mu\nu} + \frac 14 F_{\mu\nu} \frac 1{\theta^2 D^4} F_{\mu\nu}
\right)\,.
\ee
However, due to the presence of inverse covariant derivatives a non-locality with respect to the fields is present. One has to use a local form of the above action. Motivated by the work of Vilar {\etal} \cite{Vilar:2009} we have eliminate the non-local terms with the help of auxiliary fields forming  BRST-doublet structures \cite{Blaschke:2009d}:
\bea
\Gamma^{(0)} & = & \int d^4x\, \left( \frac 14 F_{\mu\nu} F_{\mu\nu} + \frac\lambda 2 ( B_{\mu\nu} + \bar B_{\mu\nu}) F_{\mu\nu} \right.\\
\nonumber
&& \hspace{.8cm}
\left. 
- \mu^2 \bar B_{\mu\nu}\star \widetilde D^2 D^2 B_{\mu\nu} + \mu^2 \bar \psi \widetilde D^2 D^2 \psi \right) \,.
\eea
In this localized version with the new doublet fields one has now a finite number of interaction terms, and it is clear now that one can define a finite number of normalization conditions in order to construct the tree approximation of the interacting part in the usual manner. However, the renormalizability of this model seems to be in doubt \cite{Vilar:2009,Blaschke:2009d}.

\section{Concluding remarks}

We have discussed general translation invariant models for noncommutative scalar and $U(1)$ gauge field theory. Due to the inherent noncommutative scale, the insertion of dimensionless operators of arbitrary power is possible. By postulating normalization conditions and using power counting arguments we could restrict this freedom. Of course, it is not clear from the start whether those are respected by quantum corrections.

In \cite{Blaschke:2009e}, the necessary IR damping of the gauge propagator has been implemented in a different manner than discussed in this note. The so-called soft breaking terms are modified, where additional sources to guarantee BRST invariance are necessary. This procedure is known from QCD, where the gluon propagator is modified in the IR in order to restrict the gauge fields to the first Gribov horizon which removes the residual gauge ambiguities. Here, we have only discussed an implementation which is gauge invariant without the help of additional sources.

The gauge action \eqref{extended-gauge} is reminiscent of the action proposed in the regularization scheme of higher covariant derivatives for non-Abelian Yang-Mills theories \cite{Faddeev:1980be,Martin:1994na,Asorey:1995tq}. There, the action contains an additional term 
\be
\int d^4x \frac 1{4\Lambda^2} \, (D^2F_{\mu\nu})^a (D^2F_{\mu\nu})^a\,,
\ee
where $\Lambda$ is a UV cut-off, and $a$ denotes the index of the gauge group. This yields for the following propagator:
\be
\label{cov-reg-propagator}
G_{\mu\nu}^{ab}(p) = \delta^{ab} \left(
\frac{\Lambda^4}{p^4(\Lambda^4+p^4)}(p^2 g_{\mu\nu} - p_\mu p_\nu) + \alpha\frac{\Lambda^4 p_\mu p_\nu}{p^4(\Lambda^4 + p^4)}\,,
\right)
\ee
where the parameter $\alpha$ characterizes the gauge fixing. The difference to the action proposed in \eqref{extended-gauge} is that the covariant derivatives only occur in the numerator. As a consequence, the propagator \eqref{cov-reg-propagator} does not show the desired IR damping behaviour. Furthermore, the consistent implementation of this regularization scheme for the case of commutative Yang-Mills theory is not at all straight forward \cite{Martin:1994na,Asorey:1995tq}.

\end{document}